%% file: main.tex
\documentclass[conference]{IEEEtran}
\usepackage{cite}
\usepackage{tikz}
\usepackage{verbatim}
\usepackage[normalem]{ulem}
\usepackage{listings}
\usepackage{amsmath,amssymb,amsfonts}
\usepackage{algorithmic}
\usepackage{graphicx}
\usepackage{textcomp}
\usepackage{xcolor}
\usepackage{fancyhdr}
\usepackage[hyphens]{url}
\usepackage{makecell}
\usepackage{multirow}
\usepackage[skip=0.5\baselineskip]{caption}
\usepackage{subcaption}
\usepackage{lipsum}

\newcommand*\circled[1]{\tikz[baseline=(char.base)]{
            \node[shape=circle,draw,inner sep=1pt] (char) {#1};}}

\newtheorem{definition}{Definition}

\newtheorem{invariant}{Invariant}
            
\def\BibTeX{{\rm B\kern-.05em{\sc i\kern-.025em b}\kern-.08em
    T\kern-.1667em\lower.7ex\hbox{E}\kern-.125emX}}

\title{Streamlining Integrity Tree Updates for Secure Persistent Non-Volatile Memory} 
\author{
    \IEEEauthorblockN{Alexander Freij\IEEEauthorrefmark{1}, Shougang Yuan\IEEEauthorrefmark{1}, Huiyang Zhou\IEEEauthorrefmark{1}, Yan Solihin\IEEEauthorrefmark{2}}\\
    \IEEEauthorblockA{\IEEEauthorrefmark{1}North Carolina State University
    \\\textit{\{atfreij,syuan3,hzhou\}@ncsu.edu}}
    \IEEEauthorblockA{\IEEEauthorrefmark{2}University of Central Florida
    \\\textit{Yan.Solihin@ucf.edu}}
}

\begin{document}
\maketitle
\pagestyle{plain}


\input{0abstract.tex}
\input{1intro.tex}
\input{2bkground.tex}
\input{3motivation.tex}

\input{4streamlining.tex}
\input{5design.tex}

\input{6eval.tex}
\input{7concl.tex}


\bibliographystyle{IEEEtranS}
\bibliography{refs}

\end{document}

%% file: 0abstract.tex
\label{sec:abstract}
\begin{abstract}

Emerging non-volatile main memory (NVMM) is rapidly being integrated into computer systems. However, NVMM is vulnerable to potential data remanence and replay attacks.

Established security models including split counter mode encryption and Bonsai Merkle tree (BMT) authentication have been introduced against such data integrity attacks. However, these security methods are not readily compatible with NVMM.
Recent works on secure NVMM pointed out the need for data and its metadata, including the counter, the message authentication code (MAC), and the BMT to be persisted atomically. However, memory persistency models have been overlooked for secure NVMM, which is essential for crash recoverability.

In this work, we analyze the invariants that need to be ensured in order to support crash recovery for secure NVMM. We highlight that prior research has substantially under-estimated the cost of BMT persistence and propose several optimization techniques to reduce the overhead of atomically persisting updates to BMTs.
The optimizations proposed explore the use of pipelining, out-of-order writes, and update coalescing while conforming to strict or epoch persistency models respectively.
We evaluate our work and show that our proposed optimizations significantly reduce the performance overhead of secure NVMM with crash recoverability.

\end{abstract}

%% file: 1intro.tex
 \section{Introduction}
\label{sec:intro}

Non-volatile main memory (NVMM) is coming online, offering non-volatility, good scaling potential, high density, low idle power, and byte addressability.  A recent NVMM example is Intel Optane DC Persistent Memory, providing a capacity of 3TB per socket~\cite{optaneDC}. Due to non-volatility, data may remain in main memory for a very long time even without power, exposing data to potential attackers~\cite{Chhabara2011iNVMM}. Consequently, NVMM requires memory encryption and integrity protection to match the security of DRAM (which we refer to as {\em secure NVMM}), or to provide secure enclave environment. Furthermore, it is expected that NVMM may store persistent data that must provide {\em crash recoverability}, a property where a system can always recover to a consistent memory state after a crash. Crash recoverability property offers multiple benefits, such as allowing persistent data to be kept in memory data structures instead of in files, and as a fault tolerance technique to reduce checkpointing frequency~\cite{Elnawawy2017RECOMPUTE,Alshboul2018LAZY,Shin2017PROTEUS, ArpitA2017ATOM, Izraelevitz2016ASPLOS}. Finally, some applications have emerged that need to run on secure enclave and yet require persistency and crash recovery, such as a shadow file system~\cite{guo19-sgxfilesystem}.  

Crash recovery of data with NVMM is achieved through defining and using memory persistency models. However, traditional memory persistency models do not automatically extend to secure NVMM, which incur two new requirements: \circled{1} {\em the correct plaintext value of data must be recovered}, and \circled{2} {\em data recovery must not trigger integrity verification failure}. To meet these requirements, the central question is what items must persist together, and what persist ordering constraints are there to guarantee the above crash recovery requirements?  Prior studies have not answered this question fully. Liu et al. pointed out that counters, data, and message authentication codes (MACs) must persist atomically~\cite{LIU2018HPCA}, but ignored the  Merkle Tree that provides integrity protection required to avoid effortless cryptanalysis. Awad et al. pointed out that Merkle Tree must also be persisted leaf-to-root~\cite{Awad2019TRIADNVM}, but did not answer what ordering requirements are needed for correct crash recovery, and how they are related to persistency models.

{\em The focus of this work is to comprehensively analyze the persist requirement and persist ordering requirements required for correct crash recovery of secure NVMM.} Getting this analysis right is important. Not only it affects correctness (i.e., whether the above crash recovery requirements are met), but it also affects performance overheads (i.e., accurate quantification of the actual performance overheads) and the reasoning of what performance optimizations are possible. For example, one property missed by prior work is that leaf-to-root updates of Bonsai Merkle trees (BMT) must follow persist order, otherwise crash recovery may trigger integrity verification failure at system recovery. Obeying this ordering constraint, we found that the overheads of crash recoverable strict persistency (SP) is about $20\times$ slowdown, which is more than one order of magnitude higher than previously reported slowdown.

In this paper, we analyze and derive invariants that are needed to ensure correct crash recovery (i.e., correct plaintext value is recovered and no integrity verification failure is triggerred). Then, to reduce the performance overheads, we propose performance optimizations, which we refer to as {\em persist-level parallelism}, or PLP, that comply with the invariants for strict and epoch persistency (EP) models. For SP, we found that pipelining BMT updates is an effective PLP optimization, which brings down the performance overheads from $7.2\times$ to $2.1\times$, compared to a secure processor model with write back caches but not supporting any persistency model. We then analyze EP where persist ordering within an epoch is relaxed, but enforced across epochs. Under EP, two more PLP optimizations were enabled besides pipelining: out-of-order BMT update and BMT update coalescing. These two optimizations reduce overheads to 20.2\%.

To summarize, the contributions of this paper are: 

\begin{itemize}
    \item To our knowledge, this is the first work that fully analyzes crash recovery correctness for secure NVMM, and formulates crash recovery invariants required under different persistency models. 
    \item For strict persistency, we propose a new optimization for pipelining BMT updates. 
    \item For epoch persistency, we propose two new optimizations: out-of-order BMT updates  
    and BMT update coalescing. 
    \item We point out that, due to incomplete reasoning of crash recovery, prior studies did not provide correct crash recovery and substantially underestimated its performance overheads. 
    \item An evaluation showing that our proposed PLP optimizations above significantly reduce the performance overhead of secure NVMM. 
\end{itemize}

The remainder of the paper is organized as follows. Section~\ref{sec:bkground} presents the background and related work. Section~\ref{sec:motivation} dives into the motivation for our work, Section~\ref{sec:method} details four BMT update systems, including the baseline used for evaluation and the three models proposed. Section~\ref{sec:design} discussed our hardware architecture, Section~\ref{sec:eval} evaluates our proposed update mechanisms, and Section~\ref{sec:concl} concludes this work.

%% file: 2bkground.tex
\section{Background and Related Work}
\label{sec:bkground}

\subsection{Threat Model}
\label{sub:threat}

We assume an adversary who has physical access to the memory system (NVMM and system bus), e.g. through ownership, theft, acquisition after system disposal, etc. Similar to the incidence of recovering sensitive data from improperly disposed used hard drives~\cite{Vijayan2011NJ, Roberts2003MIT}, data remanence in NVMM extends such vulnerabilities to data in memory~\cite{Chhabara2011iNVMM}. In addition, NVMMs are potentially vulnerable to replay attacks~\cite{Aura1997REPLAY} and cold boot attacks~\cite{Pan2018NVCool, Halderman2008COLDBOOT}, which allow malicious entities access to the systems. Similar to prior work~\cite{Lehman2016POISONIVY,Lehman2018MAPSUM, Awad2019TRIADNVM,SAILESHWAR2018MICRO,Awad2019CHALLENGE}, we assume that the adversary cannot read the content of on-chip resources such as registers and caches, hence the processor chip forms the trust boundary where trusted computing base (TCB) may be located. All off-chip devices, including main memory and memory bus, are considered vulnerable to both passive (snooping) and active (tampering) attacks. These assumptions are essential to secure processor architecture~\cite{Lie2000XOM, Suh2007AEGIS, Yang2003MICRO, Yan2006ISCA, Swami2018STASH,Chhabra2009SHIELDSTRAP, Fletcher2012SECURE}.

\subsection{Memory Encryption}

The goal of memory encryption is to conceal the plaintext of data written to the off-chip main memory~\cite{Zuo2018SECPM,Lee2005SECUREARCH, Swami2018ACME, Liu2015SEDURA}.
Counter mode encryption ~\cite{Yan2006ISCA,Rogers2007BMT, Palangappa2018CASTLE, Awad2019CHALLENGE} is commonly used for this purpose. 
It works by encrypting a counter to generate a pseudo one time pad (OTP) which is XORed with the plaintext (or ciphertext) to get ciphertext (or plaintext). To be secure, pads cannot be reused, and hence the counter must be incremented after each write back (for temporal uniqueness) and concatenated with address to form a seed (for spatial uniqueness). Counters may be monolithic (as in Intel SGX~\cite{Gueron2016MEE, Costan2016INTELSGX}) or split (as in Yan et al.~\cite{Yan2006ISCA,Rogers2007BMT}). Split counter co-locates a per-page major counter and many per-block minor counters on a single cache block, and each cache block is represented by the concatenation of a major and a minor counter. Due to its much lower memory overhead (1.56\% vs. 12.5\% with monolithic counter~\cite{Yan2006ISCA}), counter cache performance increases and the overall decryption overhead decreases. Hence, we assume the use of a split counter organization for the rest of the paper.

\subsection{Memory Integrity Verification}

Memory encrypted using counter mode encryption is vulnerable to a {\em counter replay attack} which allows the attacker to break the encryption~\cite{Yan2006ISCA}, hence memory integrity verification is needed not only to protect data integrity, but also to protect encryption from trivial cryptanalysis~\cite{Zou2019FAST, Rakshit2017ASSURE}. Data fetched from off-chip memory must be decrypted and its integrity verified when it arrives on chip. Early memory integrity protection relied on Merkle Tree covering the entire memory~\cite{Gassend2003HPCA}, with the root of the tree always kept securely on chip. When using counter mode encryption, Rogers et al. proposed Bonsai Merkle Tree (BMT)~\cite{Rogers2007BMT} that employs stateful MACs to protect data, leaving a much smaller and shallower tree covering only counters. A stateful MAC uses data, address, and counter as input to the MAC calculation; any modification to any MAC input or the MAC itself becomes detectable. Since it is sufficient to have one input component with freshness protection, BMT only needs to cover counters. Intel SGX adopted this observation to design a similar stateful MAC approach to construct a counter tree that combines counters and MACs~\cite{Gueron2016MEE}. Saileshwar et. al~\cite{SAILESHWAR2018MICRO} and Taassori et. al~\cite{Taassori2018VAULT} discussed optimizations that further reduce the BMT size.

\subsection{Intel SGX}

Secure enclaves, e.g. Intel SGX, were designed to provide secure execution environments for application software. By combining memory encryption and integrity protection with attestation and key sealing/management, it provides an application protection against compromised system software (OS and hypervisor). SGX Memory Encryption Engine (MEE) uses monolithic 56-bit counters instead of the more space efficient split counter. An integrity "counter" tree is used for integrity verification. A counter tree node co-locates counters with a MAC, the MAC accepts as input the counters it is co-located, and the parent counter for the block. This results in a highly interleaved but parallelizable integrity tree that covers the entire enclave memory ~\cite{Gueron2016MEE, Costan2016INTELSGX}.

\subsection{Memory Persistency}
Memory persistency is defined to allow the reasoning of crash recovery for persistent data, an important benefit offered by non-volatile main memory (NVMM)~\cite{Kolli2016DELEGATED,Kolli2016HIGHPERFPM,Pelley2014ISCA,Liu2019JANUS,Condit2009BETTERIO,Blelloch2018PARALLELPMEM,Dulloor2014SYSSWPMEM, Volos2011ASPLOS, Chakrabarti2014ATLAS, Ren2015THYNVM}. Specifically, it defines the visible ordering of loads and stores seen  by a crash recovery observer~\cite{Pelley2014ISCA, Lu2014LOOSE}. A persistency model defines when a store \textit{persists} (i.e. becomes durable) with respect to other stores of the same thread, and oftentimes coupled with memory consistency models to ensure visibility to other threads. 

The most conservative model, {\em strict persistency} (SP) requires that persists follow the sequential program order of stores~\cite{Pelley2014ISCA}. While providing simple reasoning, SP does not allow any overlapping or reordering of persists, limiting optimization opportunities in the system and incurring high performance overheads. More relaxed persistency models include {\em epoch persistency} (EP) and {\em buffered epoch persistency} (BEP)~\cite{Pelley2014ISCA}, as well as {\em lazy persistency}~\cite{Alshboul2018LAZY}. With EP (or BEP), programmers define regions of code that form {\em epochs}~\cite{Gogte2018SYNCFREEPM, Kolli2017LLP}. Persists within an epoch can be reordered and overlapped, but persists across epochs are strictly ordered using persist barriers, which enforce that persists in an older epoch must complete prior to the execution (or completion) of any persist from a younger epoch. 
On top of a persistency model, crash recovery often requires the programmer to define atomic durable code regions~\cite{Nalli2017WHISPER,Rudoff2017PMEMPROG,Shin2017ISCA,Zhang2019PANGOLIN,Coburn2011NVHEAPS,Dulloor2014SYSSWPMEM}.

\subsection{Secure NVMM for Crash Recovery} 

Data remanence vulnerability exists with DRAM as data may persist for weeks under very low temperature~\cite{Halderman2008COLDBOOT,Pan2018NVCool}. The vulnerability is much worse with NVM since data is retained for years, hence self-encrypting memory has been proposed~\cite{Chhabara2011iNVMM, Young2016DEUCE,Zuo2018SECPM}. However, NVM will likely host persistent data that requires supporting crash recovery, which then requires integrating memory encryption and integrity verification with memory persistency. This has been explored only recently. Swami et. al \cite{Swami2018ARSENAL} proposed co-locating data, counters, and MAC, to make it easier to atomically persist them together. Liu et al.~\cite{LIU2018HPCA} proposed a similar approach, plus an alternative approach of using the memory controller (MC) as a gathering point for atomic persistence. While necessary, these studies ignored  Merkle Tree integrity verification required for correct crash recovery. Awad et al.~\cite{Awad2019TRIADNVM} looked at persisting data, counters, and BMT, but ignored ordering requirements needed for correct crash recovery, and persistency models that are relevant for the ordering. Zuo et. al~\cite{Zuo2019SUPERMEM} proposed coalescing counter for persisting counter cache data, but did not acknowledge additional requirements for counter integrity verification.

%% file: 3motivation.tex
\section{Correctness of Crash Recovery}
\label{sec:motivation}

In this section, we formulate the invariants that need to be ensured in order to support crash recovery for secure NVMM. The system we assume is one with volatile on-chip caches, with the persistent domain including the NVMM and the write pending queue (WPQ) inside the MC. Our analysis focuses on a system with counter-mode memory encryption along with MAC and BMT memory integrity verification. Counters, MACs, and BMT nodes are cacheable and can be lost with the loss of power, except the BMT root which is always stored persistently on chip. We will discuss Intel SGX MEE later in the paper.

Suppose that plaintext $P$ at address $A$ is encrypted using counter $\gamma$ and private key {\em K} to yield ciphertext $C$, i.e., $C = E_K(P,A,\gamma)$ and necessarily the decryption follows $P = D_K(C,A,\gamma)$. Suppose also that $M$ represents a message authentication code for C, i.e., $M = MAC_K(C,A,\gamma)$. Finally suppose that BMT covers all counters and has a root $R$. We define BMT update path as follows: 

\begin{definition}
{\bf BMT update path} is the path of nodes from a leaf node (i.e., one encryption page) to the root of BMT.  
\end{definition}

\begin{figure}[htbp]
 	\centering
	\includegraphics[width=\columnwidth,scale=0.5]{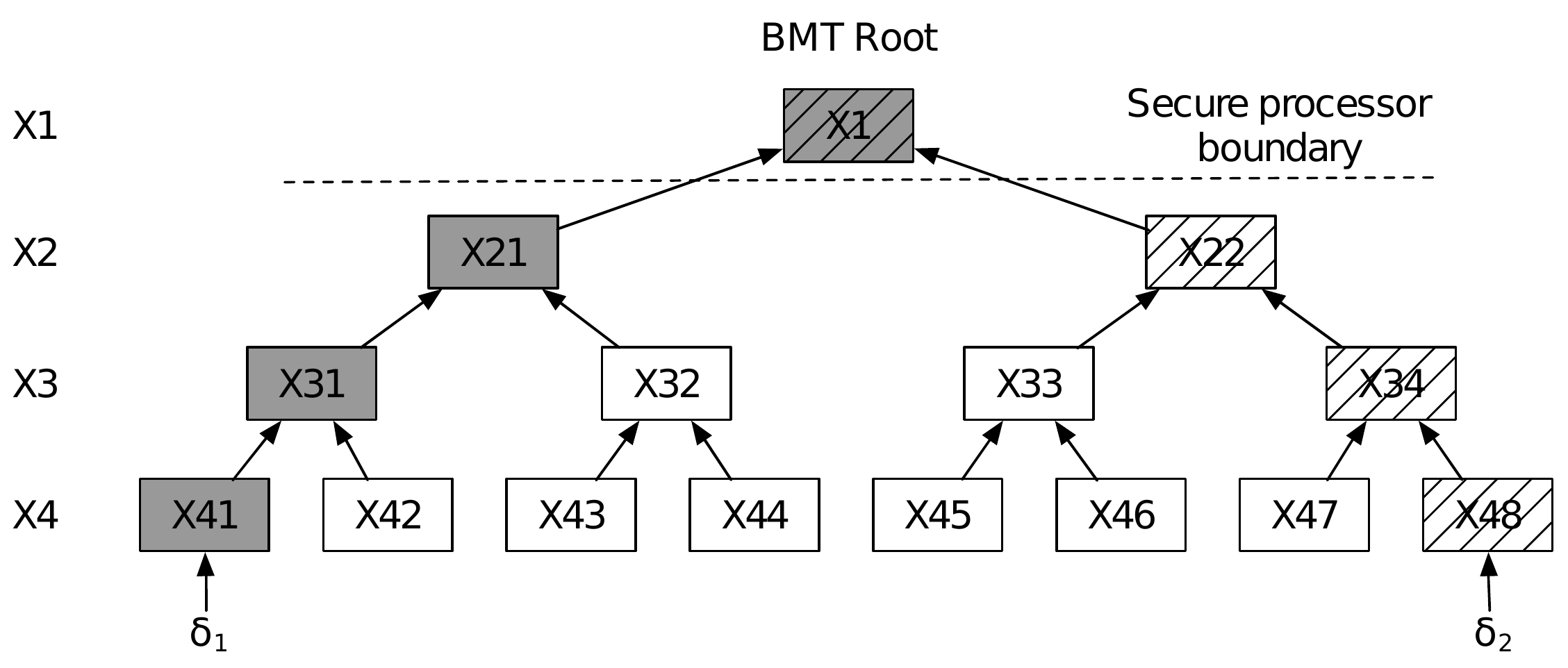}
	\caption{An example showing two BMT updates with their update paths. Persist $\delta_1$'s path is shown in grey (X41, X31, X21, X1) while $\delta_2$'s update path is striped (X48, X34, X22, X1).}
	\label{fig:baseline}
\end{figure}

Figure~\ref{fig:baseline} shows an example with two persists that generate updates to the BMT. Update $\delta_1$ is represented by nodes shown in grey while update $\delta_2$ is shown with stripes. Note that all update paths necessarily intersect at the root but the intersection can also happen earlier. 

\begin{definition}
{\bf Common Ancestors} of two persists are nodes in the BMT tree that appear in the BMT update paths of both persists. The {\bf Least Common Ancestor} (LCA) is a common ancestor that is at the lowest-to-leaf level compared to all other common ancestor nodes. 
\end{definition}

In the example in Figure~\ref{fig:baseline}, the common ancestor consists of only the BMT root, hence the BMT root is also the LCA. However, if another persist causes an update at node $X46$, then this update and $\delta_2$ share $X22$ and $X1$ as common ancestors, with $X22$ being the LCA. 

We also define a memory tuple as a collection of items that are needed to crash recover a datum: 

\begin{definition}
Secure memory transforms an on-chip plaintext data $P$ at block address $A$ to a {\bf memory tuple} of  $(C,\gamma,M,R)$ when data is persisted to main memory, and vice versa when persisted data is read from main memory. 
\end{definition}

The memory tuple represents the totality of transformation of a block when it is written back (out of the last level cache or LLC) to off-chip memory, and we claim that each tuple item must be available in order to recover data correctly, and failure to persist any item(s) in the tuple results in a crash recovery problem: 

\begin{invariant} 
{\bf Crash Recovery Tuple Invariant}. 
In a secure memory with counter-mode encryption and MAC/BMT verification, in order to recover a datum $P$ that was persisted in memory, its entire memory tuple $(C,\gamma,M,R)$ must have been persisted as well. 
\label{inv1}
\end{invariant}

To illustrate this, suppose that a plaintext value $P_o$ is changed to a new value $P_n$. The memory tuple for the block then must change from $(C_o,\gamma_o,M_o,R_o)$ to $(C_n,\gamma_n,M_n,R_n)$. If some tuple item was not persisted, for example $M_n$, post-crash, $(C_n,\gamma_n,M_o,R_n)$ is recovered. In this case, the correct plaintext is recovered but MAC verification fails because the old MAC ($M_o$) fetched from memory mismatches with $MAC_K(C_n,A,\gamma_n)$. If instead $\gamma_n$ was not persisted, since  $P_n \neq D_K(C_n,A,\gamma_o)$, the correct plaintext is not recovered. Not only that, since $\gamma_o$ is input to MAC and BMT verification, both verification mechanisms fail as well. Table~\ref{tb:cases} lists the outcomes of not persisting one or more of the memory tuple.

\begin{table}[htbp]
\centering
\caption{Recovery failure cases due to persist failure}
\label{tb:cases}
\small
\begin{tabular}{|c|c|c|c|l|} \hline
$C$ & $\gamma$ & $M$ & $R$ & Outcome \\ \hline \hline
$\surd$ & $\surd$ & $\surd$ & $\times$ & BMT (verification) failure \\ \hline
$\surd$ & $\surd$ & $\times$ & $\surd$ & MAC (verification) failure \\ \hline
$\surd$ & $\times$ & $\surd$ & $\surd$ & Wrong plaintext, BMT\&MAC failure \\ \hline
$\times$ & $\surd$ & $\surd$ & $\surd$ & Wrong plaintext, MAC failure \\ \hline
\end{tabular}
\end{table}

Note that the crash tuple invariant (Invariant~\ref{inv1}) specifies the necessary and sufficient condition for recovering data post crash. It does not specify exactly "when" tuple items must be persisted with respect to the data persist; this depends on the crash recovery expectation of the program and the persistency model being assumed. 

So far we have discussed the crash recovery correctness for a single data persist. To support crash recovery, programmers must reason about not just a single persist, but multiple persists and the relative ordering between them. In this case, we assume that if there is possibility that the crash recovery observer reads the persistent memory state between two persists, then the two persists must be ordered. Now suppose that there are two ordered persistent stores (persists) $\alpha_1$ and $\alpha_2$ to the different blocks. For the memory tuples of these different blocks, it is possible that these blocks may modify the same counter block, the same MAC block, and definitely the same BMT root. If the persist order of memory tuples is not followed, recoverability is problematic. For example, suppose that $\alpha_1 \rightarrow \alpha_2$ but $R_2 \rightarrow R_1$, which means that the BMT root is updated by the second persist before by the first persist. If a crash occurs prior to either of them or after both of them, recoverability is not jeopardized. But at other points, recovery can fail. For example, suppose that a crash occurs after $\alpha_1$ and $R_2$ persist but before $\alpha_2$ and $R_1$ persist. Post crash, BMT verification failure occurs due to the root not reflecting the persist of $\alpha_1$. In other words:

\begin{invariant}
{\bf Persist Order Invariant}. Suppose that $\alpha_1$ happens before $\alpha_2$ in program order. If the crash recovery observer may read out the persistent state between $\alpha_1$ and $\alpha_2$, then $\alpha_2$ must follow $\alpha_1$ in persist order, i.e. $\alpha_1 \rightarrow \alpha_2$. If $\alpha_1 \rightarrow \alpha_2$ in persist order, then for correct crash recovery, the following must hold: $(C_1,\gamma_1,M_1,R_1) \rightarrow (C_2,\gamma_2,M_2,R_2)$ in persist order, i.e. the persist order of each respective memory tuple items must follow the order of data persists. 
\label{inv2}
\end{invariant}

Note that the persist order depends on the persistency models that are assumed. For SP, every persist is ordered with respect to others. Hence, Invariant~\ref{inv2} applies to each pair of persists. However, for an EP model, persists are ordered or Invariant~\ref{inv2} applies only if they are from different epochs. Persists from the same epoch are unordered, which gives a rise to optimization opportunities that we will discuss in Section~\ref{sec:method}. 

The key consequence of Invariant~\ref{inv2} is that persist ordering imposes a very high cost that scales with the size of BMT. Upon eviction of a block from LLC, the data, its counter, and MAC are sent to the MC, but there they must wait until BMT update from leaf reaches the root, before the persist can be considered successful. For example, for a full 8-ary BMT constructed for an 8TB NVMM system would have a tree height of 12, meaning that for an atomic writeback of security metadata, the change to leaf nodes must traverse the 12 levels of the BMT to persist the BMT root, prior to the next persist. Assuming a hash latency of 80 processor cycles~\cite{Lehman2016POISONIVY}, this adds up to 960 processor cycles for one memory update!

%% file: 4streamlining.tex
\section{Streamlining BMT Updates}
\label{sec:method}

In this section, we explore how BMT update performance due to persists can be improved. Performance optimization techniques that are possible depend on \circled{1} no violation against invariants discussed in the previous section, and \circled{2} the persistency model that is assumed. We collectively refer to the key methods as persist-level parallelism (PLP): pipelining, out-of-order updates, and coalescing. 

\subsection{Strict Persistency}

\subsubsection{Baseline Atomic Persist Mechanism}
\label{sec:base}

Following Invariant~\ref{inv1}, for each memory update, we need to ensure that all memory tuple components also persist. Due to the write-back cache, the eviction order of dirty blocks may be different from the program order. Therefore, with SP, one way to satisfy the invariant is to atomically persist the tuple generated by each store, which results in write-through cache behavior. To achieve this, we devise a {\em 2-step persist} (2SP) mechanism.  Similar to~\cite{LIU2018HPCA}, 2SP relies on the WPQ of the MC as persist gathering point. 2SP consists of two steps: the first step involves holding and locking persist memory tuple components in the  WPQ (while flagged as incomplete), while the second step flags the completion of the persist and releases tuple components to memory. A persist is marked completed when the WPQ receives its updated counter, MAC, and acknowledgement that the BMT root has been updated. Once completed, the blocks are allowed to drain from the WPQ to the NVMM. On power failure, any incomplete flagged blocks are considered not persisted and invalidated. Since the persistence of the counter and MAC is straightforward and not expensive, we will focus the rest of the discussion on the expensive BMT update. 

\begin{figure}[hbt!]
 	\centering
	\includegraphics[width=\columnwidth,scale=0.4]{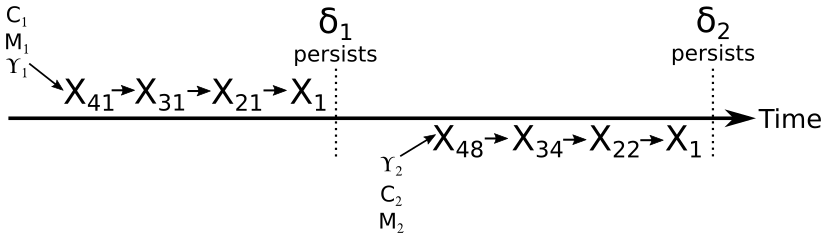}
	\caption{The timeline of two data persists and their memory tuple persists.}
	\label{fig:atomic}
\end{figure}

To illustrate the mechanism, suppose that two persists are initiated, as shown in Figure~\ref{fig:baseline}. Figure~\ref{fig:atomic} shows the sequence of persists of memory tuples due to the two persists, in the baseline persist mechanism. For persist $\delta_1$, ciphertext $C_1$, counter $\gamma_1$, MAC $M_1$ are persisted. A new value of counter $\gamma_1$ is needed for the BMT update path starting from leaf of BMT $X41$, which in turn is needed to update BMT node $X31$, and so on, until BMT root $X1$ is updated. When ciphertext $C_1$, counter $\gamma_1$, and MAC $M_1$ are completed and BMT root is updated, $\delta_1$ is considered completed, after which persist $\delta_2$ can commence. It is clear that even though intermediate nodes in the BMT update path do not need to persist (only the leaves and root need to persist), the critical path is due to their sequential updates.

\subsubsection{PLP Mechanism 1: Pipelining BMT Updates}

\begin{figure}[hbt!]
 	\centering
	\includegraphics[scale=0.3]{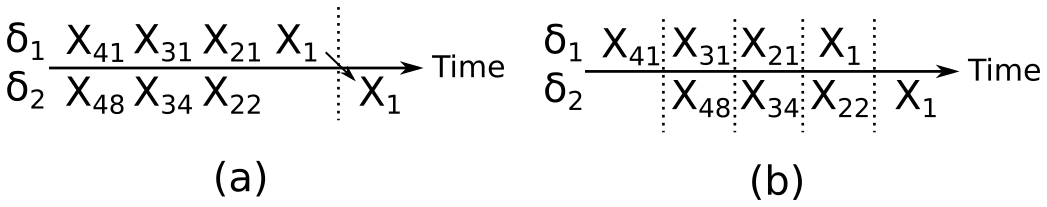}
	\caption{The timeline of (a) out-of-order BMT updates with in-order BMT root updates, and (b) pipelined updates with in-order common ancestor (including BMT root) updates.}
	\label{fig:pipelined}
\end{figure}

While the baseline persist mechanism described in Section~\ref{sec:base} is correct, it suffers from high overheads. Each node in the BMT update path must wait until the previous node has been calculated. In order to improve this situation, recall that the Persist Order Invariant (Invariant~\ref{inv2}) only requires that the BMT root update follows the persist order. This means that it is possible to update BMT nodes out of order, as long as the root is still updated in persist order. This is illustrated in Figure~\ref{fig:pipelined}(a), where update paths of persist $\delta_1$ and persist $\delta_2$ are updated out of order but updates to BMT root are kept in persist order. 

While out of order non-root updates are best for performance, it is difficult to avoid write-after-write (WAW) hazards if two persists' BMT update paths intersect at more than just the BMT root. To avoid WAW without much complexity, we design a more restrictive version of the optimization, namely {\em pipelined BMT update}. With a pipelined update, a younger persist is allowed to update a certain level of BMT only when an older persist has completed its update of the same level BMT node. This is illustrated in Figure~\ref{fig:pipelined}(b). The pipelined update optimization ensures that if two persists have common ancestor nodes, they will still be updated in persist order. 

Note that as the memory grows bigger, the BMT will have more levels and hence more pipeline stages. Thus, one attractive feature of pipelined BMT updates is that with larger memories, the degree of PLP increases and pipelined BMT updates becomes even more effective versus non-pipelined updates.

\subsection{Epoch Persistency}
\label{sec:epoch}

With EP, two persists in the same epochs do not have persist ordering constraints; persists only need to be ordered across separate epochs. This fact allows the write-back cache to reduce the write traffic and also gives us opportunities to optimize BMT updates. We make a stronger assumption on EP compared to that in literature: Nalli et. al~\cite{Nalli2017WHISPER} assert that 75\% of epochs update one 64B cache line, where we assume a minimum of one store per epoch. Specifically, we assume that crash recovery does not depend on the transient persistent state within an epoch while an epoch is executing. Instead, crash recovery depends only on the persistent state at an epoch boundary. This assumption requires that any actions performed by an epoch that were not completely persisted prior to crash must be re-executable. This assumption is reasonable, because epochs are usually components of a durable transaction, and durable transactions can be re-executed if they fail. 

\subsubsection{PLP Mechanism 2: Out-of-Order BMT Updates}

Invariant~\ref{inv2} applies to two persists that are ordered, i.e. in EP, they belong to two different epochs. It does not specify how to treat two persists that are not ordered, such as those belonging to the same epoch. The question then arises whether two unordered persists can be performed out of order (OOO), and if so, to what extent and whether there are any constraints that need to be observed. 

\begin{figure}[hbt!]
 	\centering
	\includegraphics[scale=0.4]{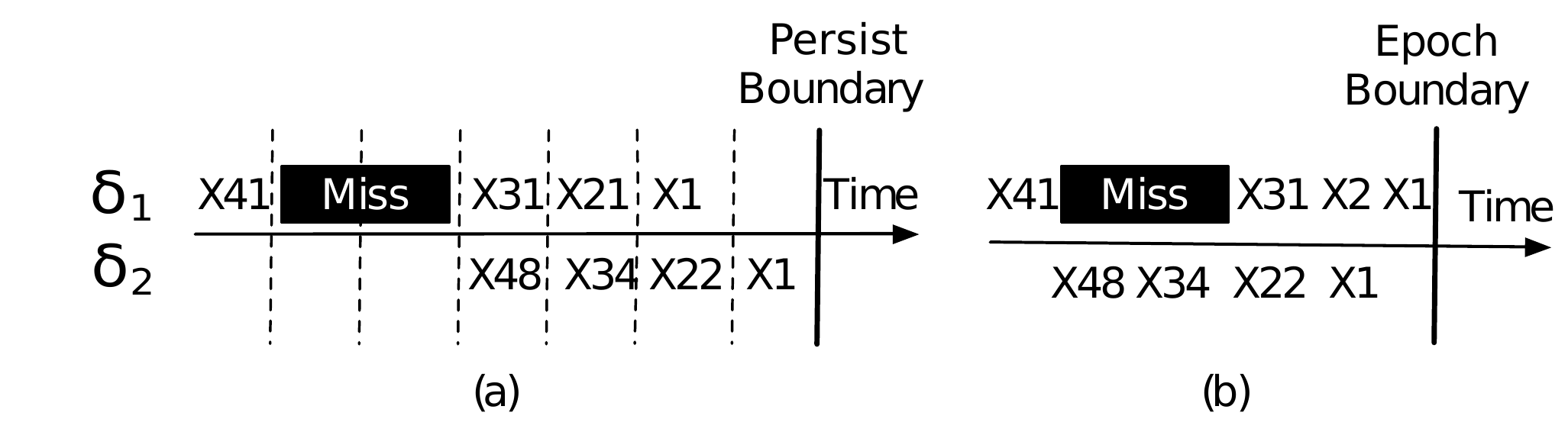}
	\caption{The timeline of two data persists with (a) in-order pipelining and (b) out of order updates.}
	\label{fig:ooo}
\end{figure}

Before discussing them further, let us first discuss the potential benefit of OOO. OOO BMT updates have a much better performance potential than (in-order) pipelining for two reasons. First, it can hide the BMT cache miss latency as illustrated in Figure~\ref{fig:ooo}. Figure~\ref{fig:ooo}(a) shows a case where persist $\delta_1$ is attempting to update the BMT, but suffers a cache miss on BMT node $X41$. This introduces bubbles in the in-order BMT update pipeline, and persist $\delta_2$ is consequently delayed, therefore it cannot update $X48$ until $X41$ is updated. Figure~\ref{fig:ooo}(b) illustrates that with OOO, both updates can occur in parallel, with $\delta_2$ not being delayed by the cache miss that $\delta_1$ must wait for. Therefore, OOO can achieve a higher degree of PLP compared to in-order pipelining. Second, OOO BMT updates enable us to use pipelined MAC units to improve the throughput. The in-order BMT update pipeline has the same number of stages as the levels in the BMT and there is at most one update at each level. Therefore, the  throughput of pipelined BMT is limited to one BMT update per $X$ cycles, where $X$ is the MAC latency. In contrast, with OOO, a BMT update can start at every cycle, thereby increasing the throughput to one BMT update per cycle. 

Regarding correctness of OOO execution of persists from the same epoch, a concern arises that there may be a write after write (WAW) hazard in the case where two persists have their BMT update paths intersecting at not just the BMT root. The hierarchical nature of BMT dictates that if two BMT update paths intersect, the intersection representing common ancestors manifests as common suffix in the paths, starting from the lowest common ancestor (LCA) node, and then continuing to the LCA's parent, grandparent, etc. until the BMT root. Does updating common ancestor nodes out of order trigger a WAW hazard? We assert that they do not. 

In order to prove it, we note that different blocks will cause different counters to be updated. Let us denote the old counter values as $\gamma_{1o}$ and $\gamma_{2o}$ and the new values as $\gamma_{1n}$ and $\gamma_{2n}$. The counters correspond to either one BMT leaf node (if the counters are co-located in a block) or two BMT leaf nodes (if the counters are not co-located in a block). In the former, the leaf node is the LCA, while in the latter the LCA is further up the tree. Suppose that persist $\delta_1$ updates the LCA before $\delta_2$. Then, at the end of the LCA update for both persists, the LCA value is $MAC_K(\gamma_{1n}, \gamma_{2n}, \ldots)$. If instead $\delta_2$ updates the LCA before $\delta_1$, the LCA value is also $MAC_K(\gamma_{1n}, \gamma_{2n}, \ldots)$, which is unchanged. Therefore, the final LCA value is the same, and hence the BMT root is also the same. The intermediate LCA value is different when $\delta_1$ or $\delta_2$ update the LCA first. However, in EP, the crash recovery observer does not expect a particular persist order for two persists in the same epoch. Furthermore, Invariant~\ref{inv2} assumes that the crash recovery observer will not read the transient persistent state between the two persists.  For the latter case, $\delta_1$ and $\delta_2$ will update different parts of the LCA, hence the same proof holds. 

The epoch boundary, however, places constraints on the degree of PLP, as it acts as point of ordering; all persists in the previous epoch must complete prior to any persist in a new epoch can complete. Thus, the higher the number of persists in an epoch, the higher is its potential PLP. 
 
To handle OOO, the 2SP only needs minor modifications. When blocks belonging to persists from the same epoch are written back from the LLC, they are no longer locked in the WPQ. They are allowed to drain to persistent memory as they come. However, the WPQ retains enough state to monitor if the memory tuples of persists of the same epoch have all arrived at the WPQ or not. When they have all arrived, they are marked completed and the epoch is considered complete. On the other hand, blocks from the next future epoch are locked in the WPQ and marked incomplete, until the previous epoch has completed.

\subsubsection{PLP Mechanism 3: BMT Update Coalescing}
\label{method-coalesce}

Further analysis of BMT updates within an EP model exposes a notable scenario that enables our final optimization. BMT updates within an epoch are likely to involve substantial number of common ancestor nodes, due to spatial locality. While OOO allows updates to BMT to be overlapped and performed out of order, there are still many updates to BMT nodes that occur. These updates can be considered superfluous considering that the same node may be updated multiple times by persists from the same epoch. In our final optimization, we seek to remove superfluous BMT updates by coalescing them.

\begin{figure}[b!]
 	\centering
	\includegraphics[width=\columnwidth,scale=0.5]{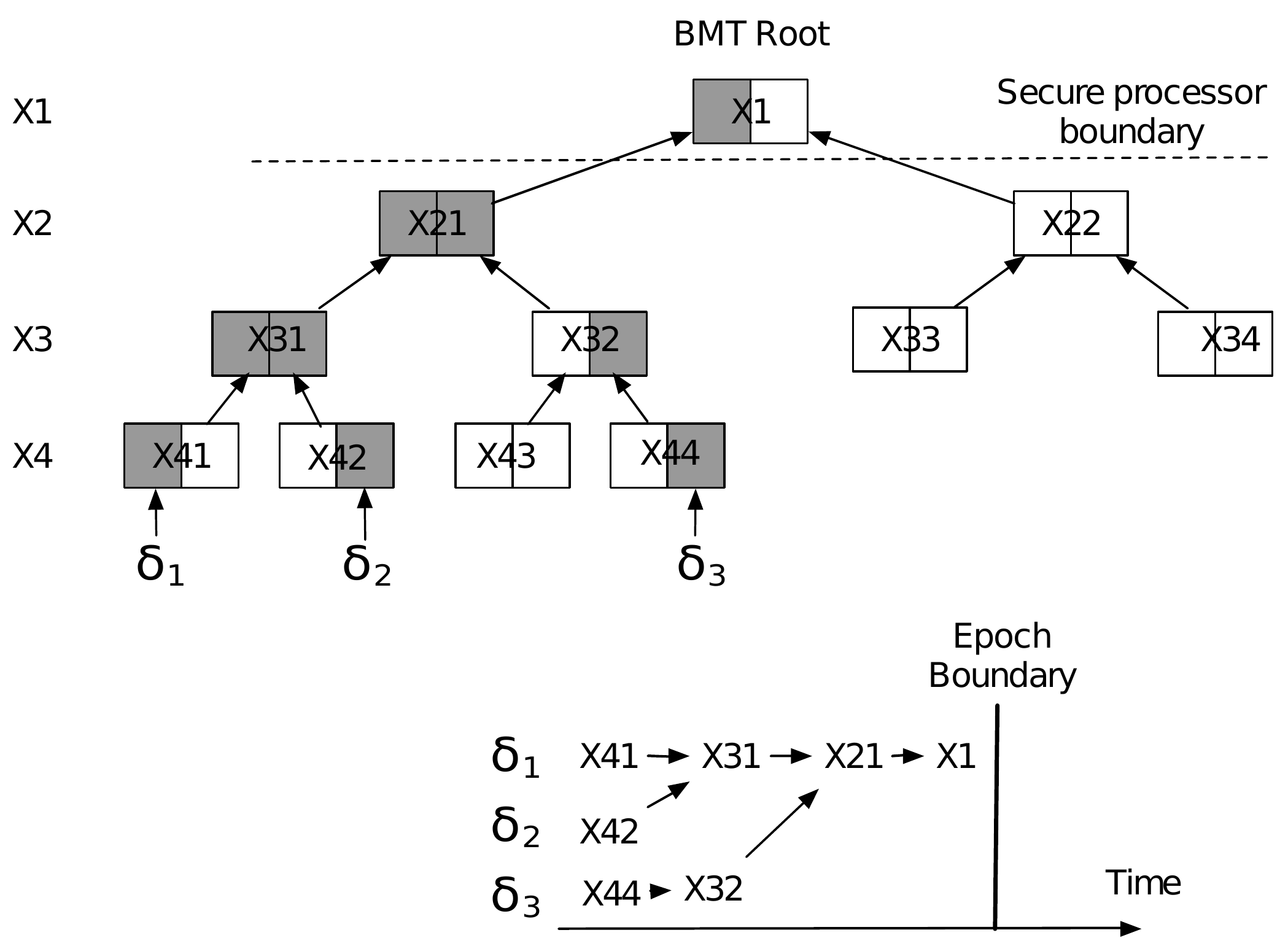}
	\caption{Example of coalescing BMT updates starting from the lowest common ancestors (LCAs) to the BMT root.}
	\label{coalesce}
\end{figure}

Figure~\ref{coalesce} illustrates the update order of OOO persists with coalescing. Without coalescing, each persist incurs updating of four BMT nodes, causing a total of 12 updates. With coalescing, persists $\delta_1$ and $\delta_2$ updates are coalesced at their LCA (node $X31$), while $\delta_3$ is coalesced at the LCA at node $X21$. As a result, there are only seven updates to the BMT, which in this example corresponds to 42\% reduction in BMT updates. Fewer updates to the BMT reduce the occupancy of the memory integrity verification engine, and hence reduces the latency and improves the throughput of the engine. Furthermore, an equally important benefit to coalescing is the number of writes. Without coalescing, the BMT root is updated three times: with coalescing, it is updated only once. 

Coalescing's effectiveness increases with spatial locality. Spatial locality results in nearby blocks being updated. In the best (and also frequent) case, blocks belonging to the same encryption page (a 4KB region) are updated within the epoch. They result in a single counter block being updated multiple times. Without coalescing, each such update generates BMT updates from leaf to root, while with coalescing, there is only one leaf-to-root update, thereby resulting in a substantial saving.  

%% file: 5design.tex
\section{Architecture Design}
\label{sec:design}
In this section, we propose architecture design to enable the PLP optimizations. As a baseline architecture, we assume a discrete counter cache~\cite{Yan2006ISCA}, BMT cache (\textit{mtcache})~\cite{Awad2018OSIRIS, Awad2019TRIADNVM}, MAC cache~\cite{Zubair2019ANUBIS}, and persist-gathering WPQ~\cite{LIU2018HPCA}. These structures suffice if an unoptimized SP model is adhered to. To support our optimizations, additional structures are introduced, specifically schedulers, to retain the persist ordering. These schedulers will contain information that enforces BMT update order by allowing or preventing writes to occur. Each optimization has its own set of conditions for allowing or preventing writes, and will be analyzed next.

\subsection{Strict Persistency Model: In-order Pipelined BMT Updates}
\label{subsec:pipeline}

To support our first PLP technique, in-order pipelined BMT updates for SP, we introduce a new structure called {\em persist tracking table} (PTT) that enforces persist ordering in a SP model.

\begin{figure}[hbt!]
 	\centering
	\includegraphics[scale=0.4]{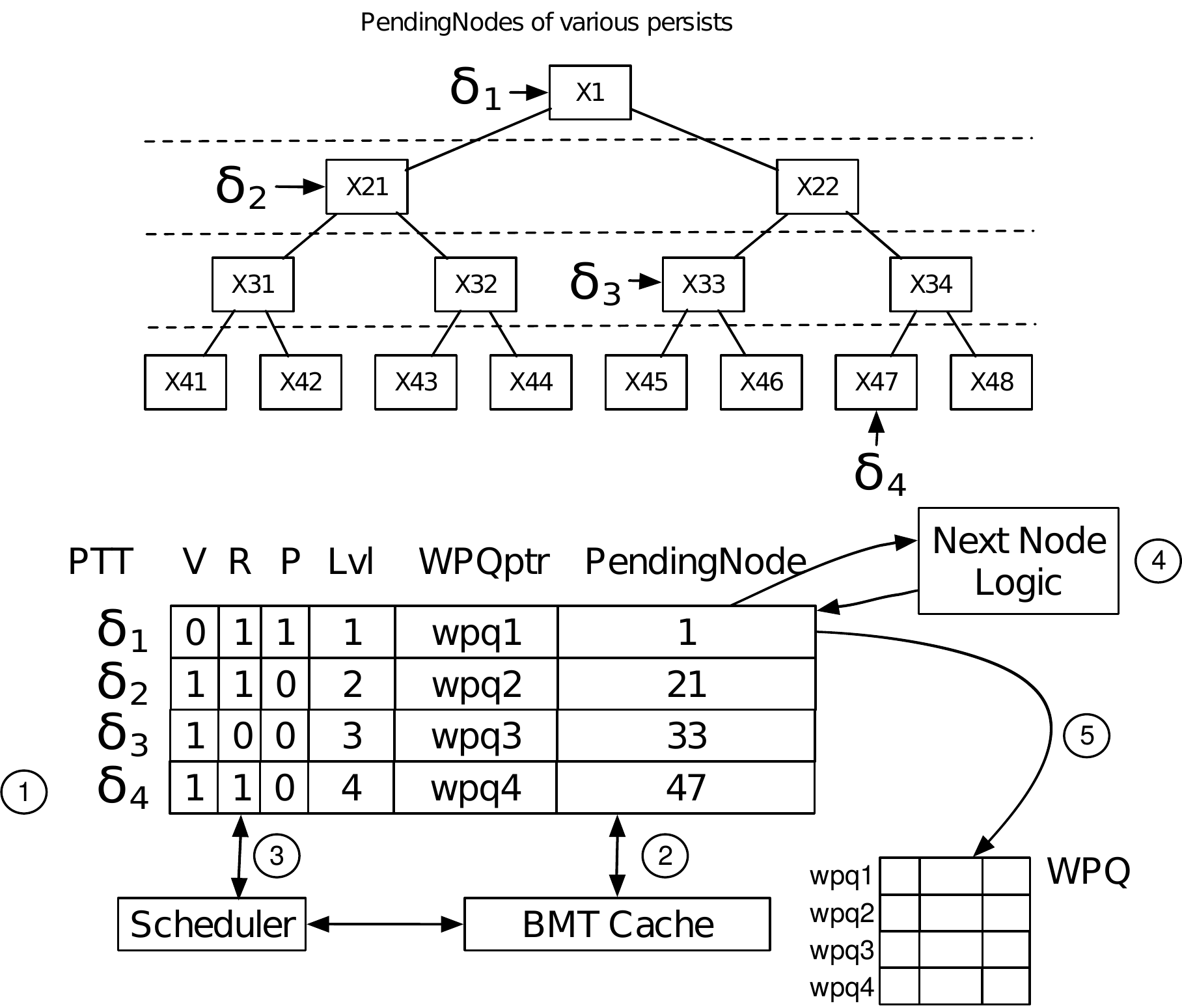}
	\caption{Example of in-order pipelined update mechanism with Persist Tracking Table (PTT) for SP.}
	\label{example-pipeline}
\end{figure}

The PTT interacts with a scheduler that also interacts with the BMT cache and the MC / WPQ. Each entry in the PTT has multiple fields (Figure~\ref{example-pipeline}). The field {\em Lvl} indicates the level of the BMT that the persist is currently updating, and is used to enforce in-order pipelining by staggering persists on different BMT levels. Figure~\ref{example-pipeline} shows an example of the PTT with four persist entries. $\delta_1$ is updating level 1 (node $X1$), while $\delta_2$ is updating level 2 (node $X21$), etc.   
The valid bit $V$ is set when the entry is created and cleared when the persist has updated the BMT root. The ready bit $R$ is set when updating the current BMT node has been completed, and cleared when the update moves on to the next node in the BMT update path. 
The PTT is managed as a circular buffer using a head and a tail pointer. The persist flag $P$ is set when the BMT root has been updated and the entry can be removed: if the head pointer points to this entry (indicating this entry being the oldest) and the \textit{P} bit is set, then BMT update is considered completed, and both the PTT entry and WPQ entry can be deallocated. The \textit{WPQptr} field points to the corresponding persist entry in the WPQ. The \textit{PendingNode} field indicates the ID/label of the node currently being updated. 

In the figure, $\delta_1$ has finished updating the BMT root hence $V=0$ and $P=1$. $\delta_2$ and $\delta$4 have updated their current nodes shown in the {\em PendingNode} fields, i.e., $X21$ for $\delta_2$ and $X47$ for $\delta$4, hence $R=1$. $\delta_3$'s $R$ bit is not set yet, either because the BMT node is not yet available for update (e.g. not found in the BMT cache/being fetched from memory), or the update has not completed (e.g., MAC is still being calculated). 

The role of the scheduler is to decide when a persist can proceed to updating the next BMT level. To illustrate the working of the scheduler, suppose a new persist request is encountered. An entry is created in the  WPQ to hold the data, counter, and MAC to persist. Concurrently, a new PTT entry is also created (Step \circled{1}), initialized to point to the corresponding WPQ entry, with the PendingNode labeled with the appropriate leaf BMT node (i.e. MAC of counter block). The valid bit is set, while the ready and persist bits are reset. In Step~\circled{2}, the BMT cache is looked up for the PendingNode. If found (BMT cache hit), a new MAC is calculated and the node updated. If not found (BMT cache miss), the node is fetched from memory, and the update commences after the node arrives from memory and is verified for integrity. Once the BMT node at the current level is updated, the $R$ bit is set. For the scheduler to allow persist entries to move on to the next BMT levels, it waits until the $R$ bits of these entries are set (Step \circled{3}), indicating completion of udpates to the current BMT levels. Once the bits are set, the scheduler wakes up the entries to move on to the next BMT levels. The PendingNode is input into the Next Node Logic to yield the ID for the next node to  update (Step \circled{4}). 

When the oldest entry ($\delta_1$) finishes updating the BMT root, the entry's $P$ bit is set and  the WPQ is notified of BMT root update completion (Step \circled{5}). Afterward, the entry occupied by $\delta_1$ can be released, the head pointer updated, and execution continues. At the WPQ, if BMT root update completion notification is received, and other tuple items are completed (data, counter, and MAC), tuple items are marked as persisted and become releasable to memory.

\subsection{Epoch Persistency Model: Out-Of-Order Pipelined BMT Updates }

\begin{figure}[hbt!]
 	\centering
    \includegraphics[scale=0.4]{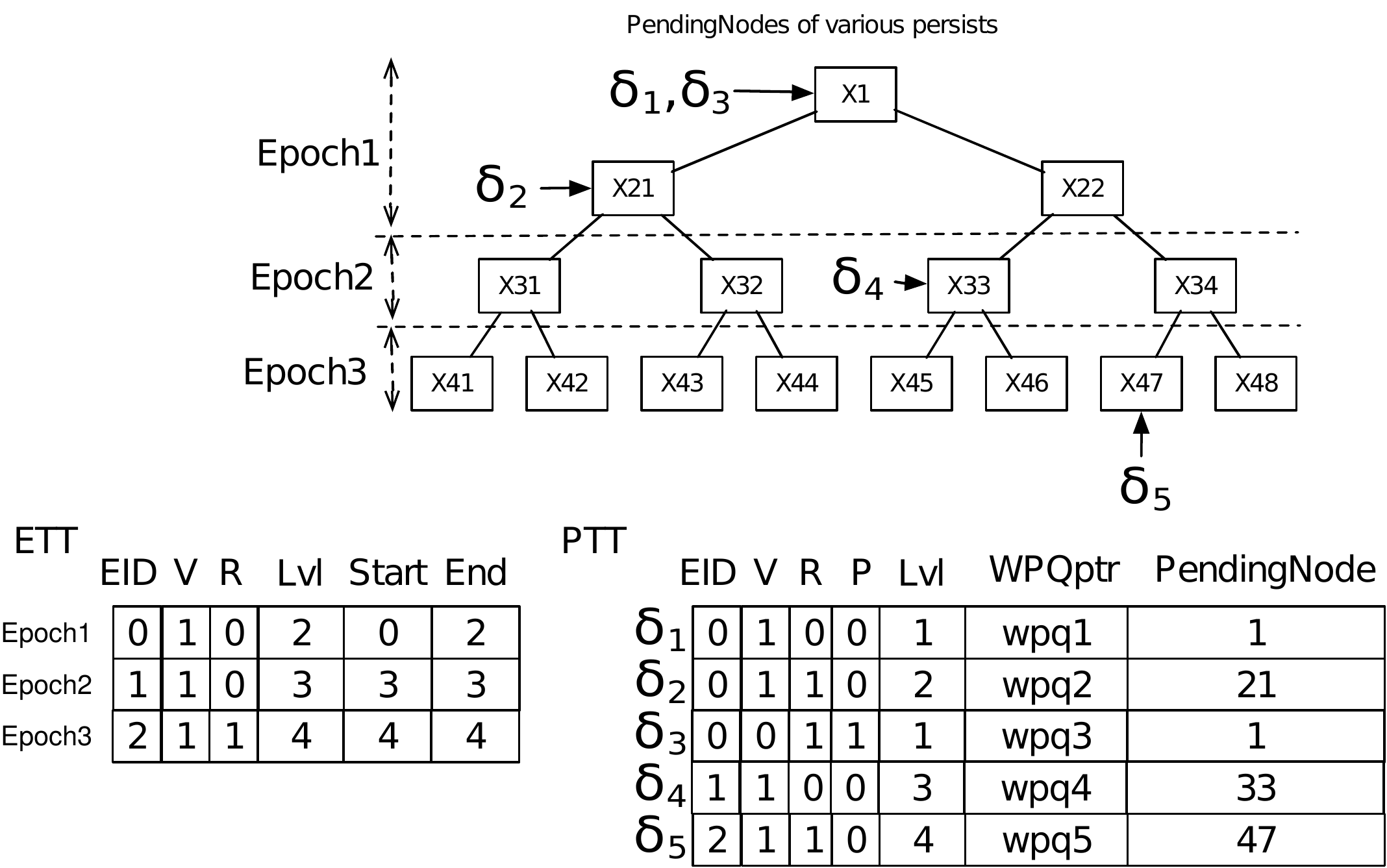}
	\caption{Proposed architecture to enable OOO BMT updates and update coalescing within an epoch as well as in-order pipelined BMT updates across different epochs.}
	\label{arch-coalesce}
\end{figure}

The previous PTT architecture is not capable of managing BMT updates with EP model with OOO updates of BMT nodes, as it enforces in-order pipelined updates. What is unique with EP is that there are two persist ordering policies: enforced ordering across epochs but not within an epoch. Thus, we split the PTT design into two tables: an {\em epoch tracking table} (ETT) to track epochs while relegating the PTT to only track persists. Furthermore, coalescing makes the PTT more sophisticated, as it must be able to calculate and track coalescing points of multiple persists. For these reasons, Figure~\ref{arch-coalesce} shows the ETT/PTT split design and also the format of the PTT entries that enable OOO updates and coalescing.

An ETT is a circular buffer maintaining the order of active epochs. An ETT entry has the following fields: \textit{EID} (epoch ID), a valid bit $V$, a ready bit $R$ (which is set when updates of all persists in the epoch are completed),  \textit{Lvl} indicating the lowest BMT level being updated by the epoch, index to the start entry at the PTT (\textit{Start}) and to the end entry at the PTT (\textit{End}).  \textit{End} is incremented (wrapped around on overflow) when a new persist from an epoch is encountered. Two special purpose registers are also added: GEC (\textit{global epoch counter}) keeps track of the next epoch ID to allocate to a new epoch, while PEC (\textit{pending epoch counter}) keeps track of the oldest active epoch being processed. In the PTT, each entry is added epoch ID (\textit{EID}) field to identify the epoch a persist belongs to. 

Figure~\ref{arch-coalesce} illustrates the tables with an example. There are a total of five persists, with the first three persists from Epoch1, while the fourth and fifth persists are from Epoch2 and Epoch3, respectively. For example, the entry for Epoch1 at the ETT has $Start=0$ and $End=2$ to indicate that PTT indices 0..2 contain information of the persists of Epoch1. $\delta_1$, $\delta_2$, and $\delta_3$ are within the same epoch, and hence they perform OOO updates on the BMT root. In the example, $\delta_3$ has updated BMT root $X1$ (hence in the PTT, $P=1$ and $V=0$), while $\delta_1$ is working on updating BMT root $X1$ (hence in the PTT, $P=0$ and $V=1$). Since $\delta_3$ has persisted, its respective entry can be released from the WPQ assuming all components of the security tuple have been received. $\delta_2$, on the other hand, has not reached BMT level 1 but has finished updating BMT node $X21$ (hence in the PTT, $R=1$. Since Epoch1 is still working on BMT level 2 node and it is the lowest level that any persist of Epoch1 is working on, in the ETT, Epoch1's $Lvl = 2$. Epoch2 and Epoch3, consisting of one persist each, are updating different nodes ($X33$ and $X47$, respectively) at different BMT levels (level 3 and 4, respectively).  

The figure illustrates that we exploit two types of parallelisms: epoch-level as well as persist-level parallelism. Within an epoch, we allow updates to occur OOO. Across epochs, we pipeline updates to the BMT in the epoch order using ETT to track and enforce correctness. The ETT mechanism for pipelining works similarly to  the PTT mechanism for pipelining for SP, but with several modifications. First, the ready bit of an epoch is set only when all its persists' ready bits are also set. The Lvl of an epoch is determined as the maximum of Lvl field of all the persists of the epoch. With this, ETT can ensure that each BMT level can only be updated by persists of a single epoch, which avoids cross-epoch WAW hazards. When all persists of an epoch's are completed within the level(s) that are recorded, an epoch's $R$ bit is set. When all epochs' $R$ bits are set, the epoch-level scheduler is invoked to advance the epochs to the next levels. If an epoch is at level 1 and its completed, the entry can then be deallocated. 

Scheduling at the PTT is also modified. In SP, persists update the BMT in a pipelined lockstep fashion. With EP, the persist's EID is used to check which level the persist is authorized to update. In the example in the figure, $\delta_5$ cannot advance to level 3 because Epoch3 is only authorized to update level 4 of the BMT. Apart from epoch-level restriction, each persist can advance to the next level independently of other persists. Hence, assuming the level is authorized, persist-level scheduler allows a persist to advance to the next level whenever $R=1$ for the persist.

\subsection{Epoch Persistency Model: Coalescing BMT Updates}

To coalesce updates within an epoch, we first need to find the common ancestors. We adopt a BMT node labeling scheme based on the previous  work~\cite{Gassend2003HPCA}. A unique label is assigned to each BMT node starting from 0 for the BMT root. To find the parent of each BMT node, we subtract one from the label of current node and divided by the arity of the BMT to get the label of its parent. Then we can round this process down until the label 0 to get a list of all its ancestors. The least common ancestor (LCA) between two leaf nodes can be found from the longest prefix match between the two ancestor lists. 

Next, we need to decide where to coalesce and how to determine which persists are coalesced together. Consider that it is likely that two persists from the same epoch will share many BMT nodes that are common. Coalescing can occur at any such node. However, the closer to leaf the common ancestor node is, the more effective coalescing become as more updates are eliminated. Therefore, an important principle for update coalescing is to coalesce at LCA whenever possible. The optimal coalescing occurs when the minimum number of updates is achieved. It requires each persist to be compared to every other persist in an epoch, and each pair that has the lowest LCA combined. Then, each combined pair is compared against every other BMT node or pair, and recombined, etc. 
However, this iterative approach is too costly for hardware implementation. Instead, we opt for paired coalescing, in which we always coalesced the new persist with previous one if it has not been coalesced with other persists.

\subsection{Streamlining Counter Tree Updates in Intel SGX}

Intel SGX utilizes a "counter tree" to verify memory integrity. Similar to BMT, the counter tree does not cover data because it assumes stateful MAC that protects against spoofing and splicing. The counter tree protects both the integrity and freshness of counters. However, unlike BMT, a counter tree requires the parent counter value to compute the MAC of child counters. As a result, to enable crash recovery, the parent counter value needs to be available and correct in order to compute the correct MAC value. On a store that persists, the tree's entire path from leaf to root nodes must also be persisted, instead of just the tree root. 

Therefore, two changes are needed for crash recovery correctness. First, Invariant~\ref{inv1} redefines a memory tuple as consisting of data ciphertext, counter, MAC, and {\em all} nodes of the counter tree from leaf to root along the update path. Consequently, Invariant~\ref{inv2} expands to include all nodes in the counter tree update path from leaf to root, in contrast to BMT which only requires the tree root to provide crash recovery. This leads to higher costs than BMT. For example, the number of updates that must persist for one store would scale by the height of the counter tree. Although the optimizations described for BMT can be adapted to SGX, we focus only on BMT due to the extra cost incurred by the counter tree.

%% file: 6eval.tex
\section{Evaluation}
\label{sec:eval}

We use the cycle-accurate simulator Gem5\cite{GEM5} to model the architecture design described in Section \ref{sec:design}. The configuration of the simulated system is presented in Table \ref{tab:config}.

\begin{table}[ht!]
\label{tab:config}
    \small
    \centering
    \begin{tabular}{|p{2cm}| p{5cm}|}
    \hline
    \multicolumn{2}{|c|}{\textbf{Processor Configuration}} \\
    \hline
    CPU & 1 core, OOO, x86\_64, 4.00GHz \\
    \hline
        L1 Cache & 8-way, 64KB, 64B block\\ 
    \hline
        L2 Cache & 512KB, 16-way, 64B block\\
    \hline
        L3 Cache & 4MB, 32-way, 64B block\\
    \hline
    \multicolumn{2}{|c|}{\textbf{Metadata Caches}} \\
    \hline
        Counter Cache & 128KB, 8-way, 64B block\\
    \hline
        MAC Cache & 128KB, 8-way, 64B block\\
    \hline
        BMT Cache & 128KB, 8-way, 9-levels tree, 64B block\\
    \hline
        MAC Latency & 40 processor cycles by default \\
    \hline
    \multicolumn{2}{|c|}{\textbf{NVM Parameters}} \\
    \hline
        \multirow{5}{*}{Memory}&8 GB DDR\_based PCM\\
        & 1200MHz clock \\& write/read queue: 128/64 entries\\&tRCD/tXAW/tBUSRT/tWR/tRFC/tCL\\&=55/50/5/150/5/12.5ns\cite{LIU2018HPCA}\\
    \hline
    \end{tabular}
    \caption{Simulation Configuration}
    \label{tab:config}
    \vspace{-0.5cm}
\end{table}

\subsection{Methodology}
\label{subsec:methodology}

Similar to previous work, \cite{Zubair2019ANUBIS,Awad2018OSIRIS, Lehman2016POISONIVY}, we utilize speculative execution for encryption/decryption mechanisms. Discrete BMT, MAC, and counter caches are implemented for all schemes discussed, with the configurations in Table \ref{tab:config}.
To enforce persist ordering, we implemented write through caches to persist each store to the MC. For pipelined BMT updates, we maintain a PTT with 64 entries. To support OOO BMT updates and coalesced BMT updates, we use a 2-entry ETT (i.e., only allow two concurrent epochs while enforcing the order between them) and a PTT with 64 entries shared by the two epochs. An {\em sfence} operation is also emulated to prevent stores from younger epoch being persisted to the memory before the stores in the elder epoch has been persisted. For our coalescing update model, we adopt a simple LCA search mechanism where two adjacent updates to the BMT can be coalesced each time, with the leading store stopping at the LCA and delegating the root update to the trailing store. 

We use 15 representative benchmarks from SPEC2006\cite{SPEC2006} to evaluate the proposed BMT write update models. All benchmarks are fast forwarded to representative regions and run with 100M instructions. The schemes we used for evaluation include:

\begin{itemize}
    \item \textbf{Secure WB model (secure\_WB):} Secure processor scheme with write-back caches, which do not support any persistency model.

    \item \textbf{Sequential update (sp):} Strict persistency model with sequential updates to the BMT.

    \item \textbf{Pipelined (pipeline):} Strict persistency model with in-order pipelined BMT updates.

    \item \textbf{Out-of-Order (o3):} Epoch persistency model with out-of-order BMT updates within each epoch but in-order across epochs.  The default epoch size, which is defined as the number of committed stores in an epoch, is 32.

    \item \textbf{Coalescing (coalescing):} Epoch persistency model with coalesced out-of-order BMT updates within each epoch but in-order across epochs.
\end{itemize}

In our sensitivity study, we vary the latency of the MAC computation, metadata cache, and epoch size to analyze their impacts.

As x86 ISA has a limited number of general purpose registers, it results in significant spills-and-refills in stack. Considering that persistent data structures mostly locate in heap or static/global region, we propose to not protect the stack region by default. The results where we evaluate full memory protection are labeled with '\_full'.

\subsection{Evaluation}
\label{subsec:evaluation}

\subsubsection{Strict Persistency}
\label{subsubsec:strict persistency}

In this experiment, we analyze the following schemes:

\begin{itemize}
    \item \textbf{sp\_full:} Atomic SP for entire memory
    \item \textbf{pipeline\_full:} pipelined write update for entire memory
    \item \textbf{sp:} Atomic SP for non-stack memory
    \item \textbf{pipeline:} Pipelined SP for non-stack memory 
\end{itemize}

\begin{figure}[hbt!]
 	\centerline{
 	\includegraphics[width=\columnwidth]{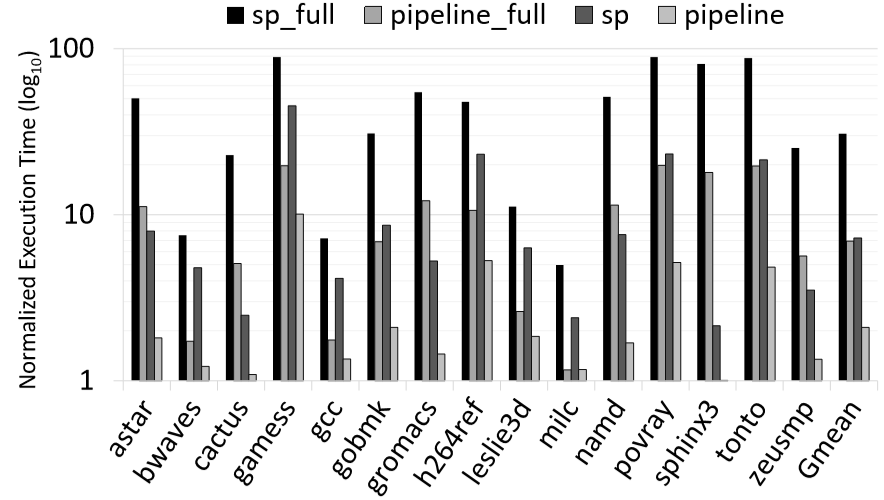}}
	\caption{Execution time of SP schemes normalized to secure\_WB model.}
	\label{strict-persistency}
\end{figure}

Figure \ref{strict-persistency} shows the execution time of these schemes normalized to the secure\_WB scheme. We can make two observations. First, SP incurs very high performance overhead, an average of $7.2\times$/$30.7\times$ for non-stack/full memory protection. The key reason is the high cost of persists. In Table~\ref{tab:number of stores}, we present the number of persists in different schemes. Take the benchmark, gamess as an example, it has 52 (non-stack) stores per kilo instructions. As each memory update needs to traverse the BMT from leaf to root and the MAC computation at each BMT level takes 40 cycles, it takes a total of 360 cycles to persist the BMT root in the 9-level BMT. As a result, the BMT update latency dominates the performance for this benchmark and we can estimate its performance in IPC (instruction per cycle) as 1000/(360*52) = 0.053, which is very close to the actual IPC, 0.054, of the SP scheme for gamess. Considering the IPC of the secure\_WB model for gamess being 2.45, the slowdown is $45.3\times$ as shown in Figure \ref{strict-persistency}. If we choose to protect the entire memory (101 stores per kilo instructions), the slowdown would further increase to $88.9\times$. Some benchmarks such as \textit{leslie3d} and \textit{bwaves} also have high numbers of persists but show lower slowdowns than gamess. The reason is that the secure\_WB model for these benchmarks have low IPC due to their high numbers of dirty-block evictions from LLC. 

Second, by overlapping the MAC computation latency, our proposed pipeline model reduces the performance overhead of SP from $7.2\times$/$30.7\times$ to $2.1\times$/$6.9\times$ for non-stack/full-memory protection. 

\begin{table}[ht!]
    \small
    \centering
    \begin{tabular}{|m{1.4cm}| m{1.3cm} |m{1.3cm} | m{1.2cm}| m{1.2cm}| }
    \hline
    Benchmark & Committed Stores & secure\_WB & sp & epoch\\
    \hline
        astar & 83.48 & 0.35 & 13.21 & 1.97 \\
    \hline
        bwaves & 100.27 & 8.70 & 61.60 & 26.47 \\
    \hline
        cactusADM & 114.59 & 1.55 & 12.35 & 5.68 \\
    \hline
        gamess & 100.72 & 0 & 51.38 & 30.433\\
    \hline
        gcc & 126.73 & 1.46 & 67.38 & 36.64\\
    \hline
        gobmk & 125.16 & 0.17 & 34.41 & 14.63 \\
    \hline
        gromacs & 105.73 & 0.04 & 9.66 & 2.69 \\
    \hline
        h264ref & 101.17 & 0 & 48.80 & 10.45\\
    \hline
        leslie3d & 108.79 & 7.78 & 58.47 & 17.58\\
    \hline
        milc & 40.18 & 2 & 13.65 & 4.10\\
    \hline
        namd & 133.10 & 0.18 & 19.66 & 2.07\\
    \hline
        povray & 150.72 & 0 & 39.23 & 11.22\\
    \hline
        sphinx3 & 184.29 & 0.10 & 4.87 & 1.04\\
    \hline
        tonto & 141.84 & 0 & 34.45 & 16.60\\
    \hline
        zeusmp & 175.87 & 1.92 & 19.87 & 4.66\\
    \hline
        average & 119.51 & 1.61 & 32.60 & 12.41\\
    \hline
    \end{tabular}
    \caption{Number of persists per kilo instructions. The numbers in 'committed stores' and 'secure\_WB' include both stack and non-stack accesses while others exclude stack accesses.}
    \label{tab:number of stores}
\end{table}

\begin{figure}[hbt!]
 	\centerline{
 	\includegraphics[width=\columnwidth]{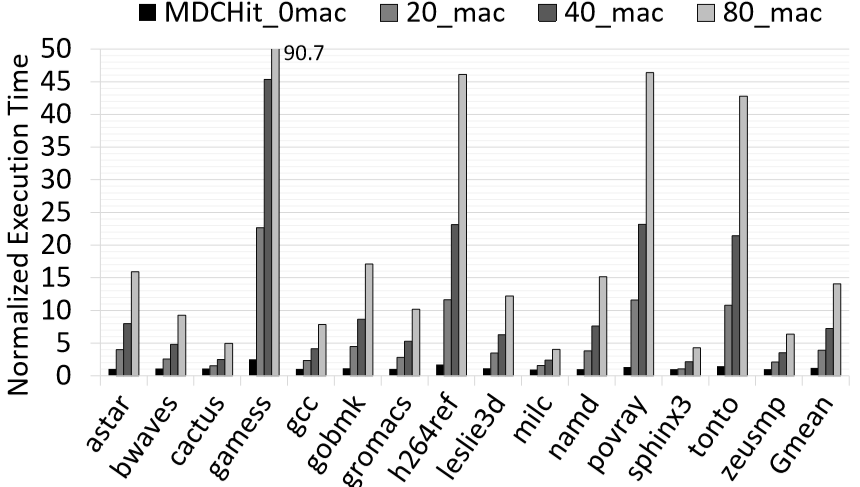}}
	\caption{Execution time of the SP scheme with different MAC latency and idealistic/realistic MDC}
	\label{MACSensitivity}
\end{figure}

To better understand the impact of MAC latency, in the next experiment, we vary the MAC latency from 0, 20, 40 and 80 cycles. We also simulate idealistic meta-data caches (MDC) to study the impact from these caches. The results are shown in Figure~\ref{MACSensitivity}. From the figure, we can see that the main performance bottleneck for SP is indeed the MAC computation latency. Even a 20-cycle MAC computation latency leads to a slowdown of $3.2\times$ on average. The MDC has negligible impact in comparison.

\subsubsection{Epoch Persistency}
\label{subsubsec:epoch persistency}
\begin{figure}[hbt!]
 	\centerline{
 	\includegraphics[width=\columnwidth,scale=0.6]{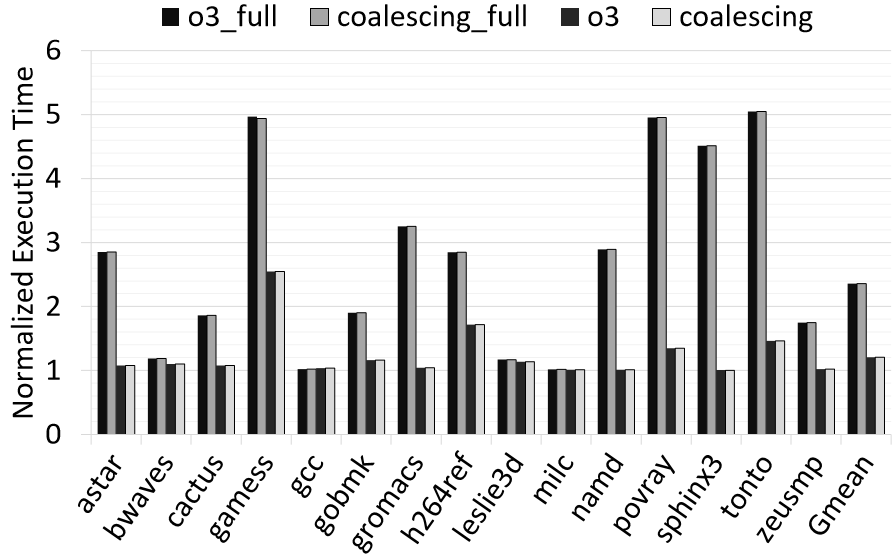}}
	\caption{Execution time of epoch persistency schemes normalized to secure\_WB model.}
	\vspace{-0.2cm}
	\label{epoch-persistency}
\end{figure}

Figure \ref{epoch-persistency} reports the performance of our epoch O3 and coalescing models with an epoch size of 32. 

Our results show that using EP scheme to protect non-stack memory, O3 and coalescing model reduces the performance slowdown to 20.7\% and 20.2\%, respectively, compared to the secure\_WB model. The performance improvements mainly come from OOO BMT updates, which enables aggressive overlapping of MAC latency. Furthermore, a large epoch also reduces the number of stores that need to be persisted if they update the same cache line. Such reduction is reported in Table~\ref{tab:number of stores}. On the other hand, if stack memory needs to be protected, the frequent updates lead to higher performance overhead, $2.42\times$ and $2.35\times$ for O3 and coalescing model, respectively. 

The results in Figure \ref{epoch-persistency} also show that coalescing has limited impact on performance. The reason is that in order to coalesce updates, the older update would wait for the younger one to reach the LCA. Therefore, the saving from coalescing is mainly the number of updates to the BMT. Our experiments show that our coalescing scheme reduces the BMT updates by 26.1\% on average.

Another interesting observation from Figure \ref{epoch-persistency} is that our optimized epoch persistency model can achieve slightly better or equal performance compared to secure\_WB model for some benchmarks like milc. The reason is that in the secure\_WB model, the evicted dirty blocks from LLC perform BMT updates sequentially rather than the OOO pipelined manner in our optimized model.

\subsubsection{Epoch size}
\label{subsubsec:epoch size}
\begin{figure}[hbt!]
 	\centerline{
 	\includegraphics[width=\columnwidth,scale=0.4]{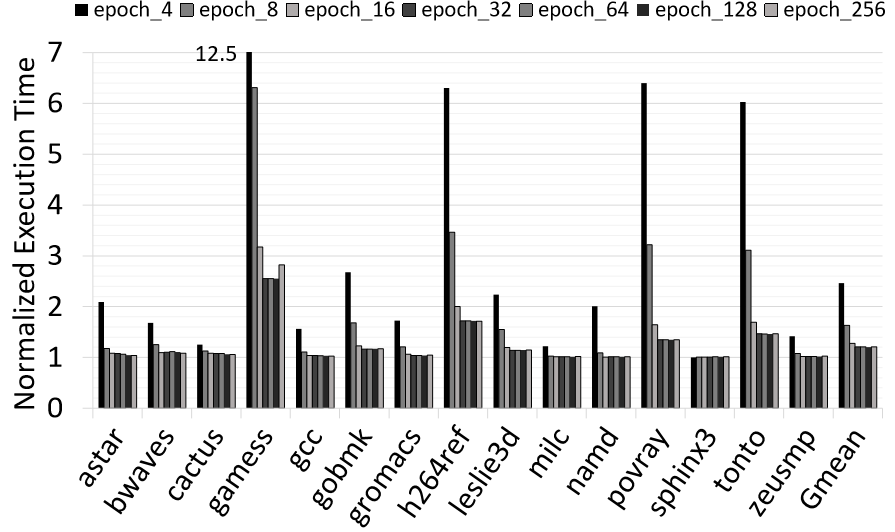}}
	\caption{Execution time of our optimized epoch persistency model with different epoch sizes normalized to secure\_WB model.}
	\vspace{-0.2cm}
	\label{epoch-size}
\end{figure}

\begin{figure}[hbt!]
 	\centerline{
 	\includegraphics[width=\columnwidth,scale=0.4]{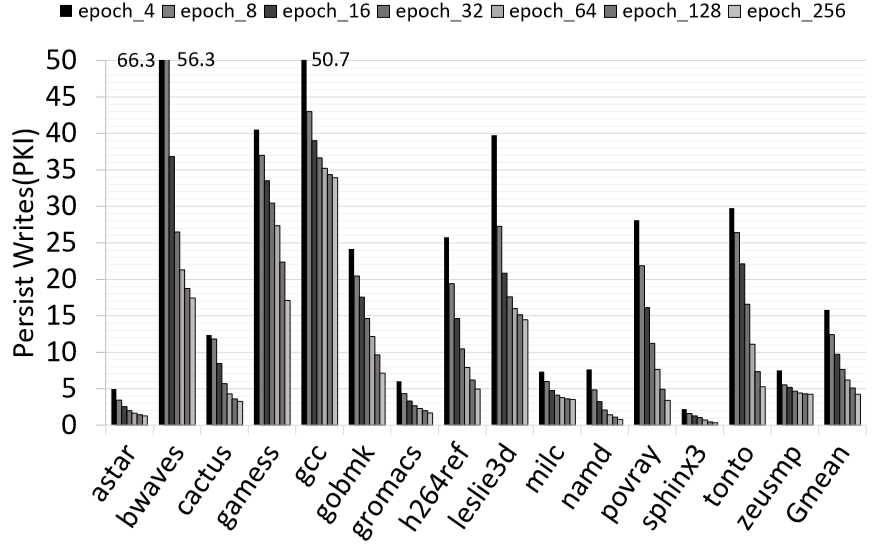}}
	\caption{The number of persists per kilo instruction for different epoch sizes.}
	\vspace{-0.2cm}
	\label{epoch-PPKI}
\end{figure}

Figure~\ref{epoch-size} shows the performance results of varying the epoch size for our optimized epoch persistency model. Besides determining the size of the PTT in our design, the epoch size has interesting performance implications. On one hand, large epochs enable higher reductions in the number of persists, i.e., making a better use of WB caches, as shown in Figure~\ref{epoch-PPKI}. On the other hand, large epochs lead to bursty memory updates at the end of each epoch. In contrast, small epochs smooth the write traffic and benefit from eager write-back \cite{DBLP:conf/micro/LeeTF00} at the cost of higher numbers of persists. In the extreme case, when the epoch size is 1, our epoch persistency model is essentially the same as the SP model. This performance tradeoff is evident in our results shown in Figure ~\ref{epoch-size}. For small epoch sizes less than 16, multiple benchmarks show high performance overhead due to the high number of persists. For the large epoch size of 256, benchmarks such as gamess, milc, and zeusmp exhibit inferior performance than that for the epoch size of 128. Based on such performance trends, we choose the epoch size of 32 due to its good performance at relatively low hardware cost.

\subsubsection{Metadata Cache Size}
\label{subsubsec:epoch size}
In this experiment, we vary the metadata cache size from 32KB to 256KB. The metadata caches include a counter cache, a MAC cache, and a BMT cache. Our results show that our persistency models are not sensitive to the metadata cache capacity and there is up to 2\% performance difference when we change the cache sizes.

%% file: 7concl.tex
\section{Conclusions}
\label{sec:concl}
Memory integrity verification and encryption are essential for implementing secure computing systems. Atomically persisting memory integrity tree roots is responsible for the majority of the overhead incurred by updating security metadata. In this work, we presented three optimizations for atomically persisting NVM Bonsai Merkle Tree roots. With a strict persistency model, our proposed pipelined update mechanisms showed an $3.4\times$ performance improvement compared to sequential updates. Within an epoch persistency model, our out-of-order root update and update coalescing mechanisms showed additional performance improvements of $2.8\times$ over sequential updates. These optimizations significantly reduce the time required to update integrity tree roots and pave the way to make secure NVMM practical.